\documentstyle[prb,aps,eqsecnum,epsf,preprint]{revtex}

\begin{document}

\tightenlines

\title{Effective Analysis of the O(N) Antiferromagnet: Low
Temperature Expansion of the Order Parameter}

\author{Christoph P. Hofmann \cite{Adr}}

\address{Institute for Theoretical Physics, University of Bern,
Sidlerstrasse 5, CH-3012 Bern, Switzerland}

\date{June 1997}

\maketitle

\begin{abstract}
\noindent
We investigate the low energy properties of Lorentz-invariant theories with a
spontaneously broken rotation symmetry O(N) $\to$ O(N--1). The leading
coefficients of the low temperature expansion for the partition function are
calculated up to and including three loops. Emphasis is put into the special
case N=3: it describes the antiferromagnet which has been extensively studied.
Our results obtained within the framework of the effective Lagrangian technique
are compared with the literature. In particular, we show that, at order $T^7$
for the heat capacity and $T^6$ for the order parameter, respectively,
logarithmic terms appear in the low temperature expansion, which have been
overlooked so far. 
\end{abstract}


\section{Introduction}
\label{Intro}

In the present paper, we investigate the low temperature properties of
spontaneously broken O(N)-symmetric theories. More precisely, we consider the
specific case where the symmetry G = O(N) of the Lagrangian is spontaneously
broken to H = O(N--1) -- we then have N--1 Goldstone bosons in the broken
phase (N $\! \geq $2). If the perturbations, which break the rotation symmetry
explicitly, are small, these excitations remain light and dominate the low
energy behavior of the system. Moreover, the Goldstone particles interact
weakly at low energies and a systematic perturbative expansion can be advised.
In particular, the partition function can be evaluated in this manner,
amounting to low temperature theorems for quantities of physical interest.

A very efficient tool to analyze the low energy structure of a system which
exhibits spontaneous symmetry breaking is provided by the effective Lagrangian
technique. The method applies to any system where the Goldstone bosons are the
only excitations without energy gap. The essential point here is that the
properties of these degrees of freedom and their mutual interaction are
strongly constrained by the symmetry inherent in the underlying model -- the
specific nature of the underlying model itself, however, is not important. For
a pedagogic outline of the effective Lagrangian method with applications to
nonrelativistic systems, the reader may consult reference \cite{Hofmann
Scattering}, which is written in a condensed matter perspective.
\cite{footnote1}

Within the framework of effective Lagrangians, the low temperature behavior of
chiral theories, where the respective groups are G = $ \mbox{SU(n)}_L
\times \mbox{SU(n)}_R $ and H = $\mbox{SU(n)}_V\,$, has been
analyzed in detail \cite{Gasser Leutwyler Temperature,Gerber Leutwyler}. In
particular, the low temperature expansion for the quark condensate, which
represents the most prominent order parameter in quantum chromodynamics (QCD),
has been evaluated to three loops. In the present work, we repeat this analysis
for Lorentz-invariant theories displaying a spontaneously broken rotation
symmetry, O(N) $\to$ O(N--1) -- the system may then be referred to as O(N)
antiferromagnet.

Apart from this general analysis, our interest is devoted to the special case
N=3: the results obtained then describe the O(3) antiferromagnet, where the
spin waves or magnons represent the corresponding Goldstone degrees of freedom.
This system has been widely studied in condensed matter physics and it is
instructive to compare our results with the findings derived within the
microscopic Heisenberg model. The lattice structure of a solid singles out
preferred directions, such that the effective Lagrangian is not invariant under
space rotations. In the case of a cubic lattice, the anisotropy, however, only
shows up at higher orders of the derivative expansion \cite{Hasenfratz
Niedermayer} -- the discrete symmetry of the lattice thus implies space
rotation symmetry. Hence, the leading order effective Lagrangian of an
antiferromagnet is Lorentz invariant \cite{Leutwyler NRD}: antiferromagnetic
spin-wave excitations exhibit relativistic kinematics, with the velocity of
light being replaced by the spin-wave velocity.

In a Lorentz-invariant theory, the following invariance theorem holds: up to a
Wess-Zumino term, the effective Lagrangian may be brought to a form which is
manifestly invariant with respect to the internal symmetry of the underlying
theory \cite{Leutwyler foundations}. The procedure of constructing the
corresponding effective Lagrangian is thus straightforward: one writes down the
most general expression consistent with Lorentz symmetry and the internal
symmetry G of the underlying model in terms of Goldstone fields $U^a(x), a = 1,
\dots$, dim(G) - dim(H) -- the effective Lagrangian then consists of a string
of terms involving an increasing number of derivatives or, equivalently,
amounts to an expansion in powers of the momentum. Moreover, the effective
method allows to systematically take into account interactions which explicitly
break the symmetry G of the underlying model, provided that they can be treated
as perturbations.\cite{footnote2}

In the particular case we are considering, the symmetry O(N) is broken by an
external field: it is convenient to collect the (N--1) Goldstone fields $U^a$
in a N-dimensional vector $U^i = (U^0,U^a)$ of unit length,
\begin{equation}
U^i(x) \, U^i(x) \, = \, 1 \, ,
\end{equation}
and to take the constant external field along the zeroth axis,
$H^i = (H,0, \dots , 0)$. The Euclidean form of the effective Lagrangian up to
and including order $p^4$ then reads \cite{Hasenfratz Leutwyler}:
\begin{eqnarray}
\label{Leff}
{\cal L}_{eff} \ = \ \mbox{$ \frac{1}{2}$} F^2 {\partial}_{\mu}
U^i{\partial}_{\mu} U^i \, - \, {\Sigma}_s H^i U^i \, - \, e_1
({\partial}_{\mu} U^i {\partial}_{\mu} U^i)^2 \, - \, e_2 \, ({\partial}_{\mu}
U^i {\partial}_{\nu} U^i)^2 \nonumber\\
+ \, k_1 \! \, \frac{{\Sigma}_s}{F^2} \,
(H^i U^i) ({\partial}_{\mu} U^k {\partial}_{\mu} U^k) \, - \, k_2 \, \!
\frac{{\Sigma}_s^2}{F^4} \, (H^i U^i)^2 \, - \, k_3 \, \!
\frac{{\Sigma}_s^2}{F^4} \, H^i \! H^i \, .
\end{eqnarray}
In the power counting scheme, the field $U(x)$ counts as a quantity of order
one. Derivatives correspond to one power of the momentum, ${\partial}_{\mu}
\propto p$, whereas the external field $H$ counts as a term of order $p^2$.
Hence, at leading order ($\propto p^2$) two coupling constants, $F$ and
${\Sigma}_s$, occur, at next-to-leading order ($\propto p^4$) we have five such
constants, $e_1, e_2, k_1, k_2, k_3$. Note that these couplings are not fixed
by symmetry -- they parametrize the physics of the underlying theory.

The effective Lagrangian method provides us with a simultaneous expansion in
powers of the momenta and of the external field. To a given order in the low
energy expansion only a finite number of coupling constants and a finite number
of graphs contribute. Consider e.g. scattering processes between Goldstone
particles. The leading contributions to the scattering amplitudes stem from the
tree graphs. Graphs involving $l$ loops are suppressed by $l$ powers of
$p^2/F^2$ and do therefore not affect the leading terms.

If the scattering amplitudes are thus needed to accuracy $p^4$, the effective
Lagrangian must be known up to and including ${\cal L}^4_{eff}$. There are two
types of contributions at this order of the low energy expansion: one-loop
graphs of ${\cal L}^2_{eff}$ and tree graphs containing one vertex from
${\cal L}^4_{eff}$. Similarly, at order $p^6$, two-loop graphs of
${\cal L}^2_{eff}$ as well as one-loop graphs involving one vertex from
${\cal L}^4_{eff}$ and tree graphs with two vertices from ${\cal L}^4_{eff}$
or one vertex from ${\cal L}^6_{eff}$ contribute etc.

It is convenient to use dimensional regularization of the loop integrals. In
this scheme, the ultraviolet divergences of the one-loop graphs of
${\cal L}^2_{eff}$ are absorbed in a renormalization of the coupling constants
which occur in ${\cal L}^4_{eff}$ -- the constants $F$ and ${\Sigma}_s$ in
${\cal L}^2_{eff}$ are not renormalized. More generally, dimensional
regularization ensures that the ultraviolet divergences which occur in the
{\it sum} of all graphs of order $p^{2n}$ are removed by a suitable
renormalization of the coupling constants in ${\cal L}^{2n}_{eff}$ -- the lower
order Lagrangians ${\cal L}^2_{eff}, \ldots, {\cal L}^{2n - 2}_{eff}$ remain
untouched.

\section{Finite Temperature}
\label{Finite T}

The effective Lagrangian method can readily be extended to finite temperature.
In the partition function, contributions of massive particles are suppressed
exponentially, such that the Goldstone bosons dominate the properties of the
system at low temperatures. Hence, the low energy theorems for scattering
amplitudes, e.g., are converted into temperature theorems for the partition
function. In the power counting rules, the role of the external momenta is
taken over by the temperature, which is treated as a small quantity of order
$p$. The interaction among the Goldstone degrees of freedom now generates
perturbations of order $p^2/F^2 \propto T^2/F^2$.

In the effective Lagrangian framework, the partition function is represented as
a Euclidean functional integral,\cite{footnote3}
\begin{equation}
\label{TempExp}
\mbox{Tr} \, [\exp(-{\cal H}/T)] \, = \, \int [{\mbox{d}}U] \,
\exp(- {\int}_{\!\!\! T} \! {\mbox{d}}^4x \, {\cal L}_{eff}) \, .
\end{equation}
The integration is performed over all field configurations which are
periodic in the Euclidean time direction, $U({\vec x}, x_4 + \beta) = 
U({\vec x}, x_4)$, with $\beta \equiv 1/T$. The low temperature expansion of
the partition function is obtained by considering the fluctuations of the field
$U$ around the ground state $V = (1, 0, \dots, 0)$, i.e. by expanding $U^0$ in
powers of $U^a$, $U^0 = \sqrt{1-U^aU^a}$. The leading contribution
(order $p^2$) contains a term quadratic in $U^a$ which describes free Goldstone
bosons of mass $M^2 = {\Sigma}_s H / F^2$. The remainder of the effective
Lagrangian is treated as a perturbation. Evaluating the Gaussian integrals in
the standard manner, one arrives at a set of Feynman rules which differ from
the conventional rules of the effective Lagrangian method only in one respect:
the periodicity condition imposed on the Goldstone field modifies the
propagator. At finite temperature, the propagator is given by
\begin{equation}
\label{ThermalPropagator}
G(x) \, = \, \sum_{n \,= \, - \infty}^{\infty} \Delta({\vec x}, x_4 + n \beta)
\, ,
\end{equation}
where $\Delta(x)$ is the Euclidean propagator at zero temperature. We restrict
ourselves to the infinite volume limit and evaluate the free energy density
$z$, defined by
\begin{equation}
\label{freeEnergyDensity}
z \ = \ - \, T \, \lim_{L \to \infty} L^{-3} \, \ln \, [\mbox{Tr}
\exp(-{\cal H}/T)] \, .
\end{equation}

Temperature thus produces remarkably little change : to obtain the partition
function, one simply restricts the manifold on which the fields are living to a
torus in Euclidean space. The effective Lagrangian remains unaffected -- the
coupling constants $F, {\Sigma}_s, e_1, \dots$ are temperature independent.

To evaluate the graphs of the effective theory, it is convenient to use
dimensional regularization, where the zero temperature propagator reads
\begin{equation}
\label{regprop}
\Delta (x) \, = \, (2 \pi)^{-d} \! \int \! \! {\mbox{d}}^d p \, e^{ipx} (M^2
\! + p^2)^{-1} \, = \, {\int}_{\!\!\! 0}^{\infty} \mbox{d} \rho \, (4 \pi
\rho)^{-d/2} \, e^{- \rho M^2 - \, x^2/{4 \rho}} \, ,
\end{equation}
We perform the calculation up to and including terms of order $p^8$ in the free
energy density of the system. To this order in the momenta, contributions to
the effective Lagrangian involving at most eight derivatives enter and the
perturbative expansion requires the evaluation of graphs containing at most
three loops.

The renormalization procedure is identical with the one used in connection with
chiral effective theories: the same graphs have to be evaluated and, up to
Clebsch-Gordan coefficients specific to the group O(N), the same ultraviolet
divergent expressions occur in the loop-integrals -- for the explicit form of
these quantities and the construction of the respective counterterms, the
reader is thus referred to reference \cite{Gerber Leutwyler}. We just mention
the fact that the contributions to the effective Lagrangian of order $p^6$ and
of order $p^8$ merely renormalize the mass of the Goldstone bosons and the
vacuum energy. At the order in the low temperature expansion we are considering
here, the values of the coupling constants occurring in these pieces of the
effective Lagrangian are therefore irrelevant and the renormalized mass takes
the form
\begin{equation}
\label{renMassH}
M_{\pi}^2 \ = \ \frac{{\Sigma}_s H}{F^2} \, + \, \frac{N\!-\!3}{32{\pi}^2}
\, \frac{({\Sigma}_s H)^2}{F^6} \ln\frac{H}{H_{M}} \; + \; {\cal O} (H^3)
\, .
\end{equation}
The logarithmic scale $H_{M}$ is determined by $k_1$ and $k_2$, i.e. by
two next-to-leading order coupling constants (order $p^4$) -- a brief
discussion can be found in the appendix.

In the limit of a zero external field, the low temperature expansion of the free
energy density is a power series of the type
\begin{equation}
\label{powerstructure}
z \; = \; \sum_{m,n \, = \, 0,1, \dots} \, c_{mn} \, {(T^2)}^m {(T^2 \,
\log T)}^n \, ,
\end{equation}
and the evaluation of the graphs of the effective theory corresponds to a
calculation of the coefficients in this series, which are pure numbers.

If the external field $H$ is different from zero, the low temperature expansion
is not a simple power series in $T$ and $\log T$. The free energy density
then involves nontrivial functions of the ratio $M_{\pi}/T$. To analyze the
behavior of the system at temperatures of the order of $M_{\pi}$, we treat both
$T$ and $M_{\pi}$ as small quantities compared to the scale of the underlying
theory,\cite{footnote4} allowing the ratio $M_{\pi}/T$
to have any value (simultaneous expansion in powers of $T$ and of $M_{\pi}$ at
{\it fixed} ratio ${M_{\pi}/T}$). The infrared singularities involving negative
powers of $T$ are thus removed by reordering, i.e. writing the series in terms
of the two variables $T$ and $M_{\pi}/T$ and ordering powers of $T$. In this
generalized sense, the low temperature expansion of the free energy density is
a power series of the form (\ref{powerstructure}) even for nonzero external
field; the symmetry breaking merely affects the coefficients $c_{mn}$ which now
become nontrivial functions of the ratio $M_{\pi}/T$.

\section{Results}
\label{Results}

In this section, we are going to discuss the low temperature properties of the
O(N) antiferromagnet, described by the effective Lagrangian (\ref{Leff}).

Since the system is homogeneous, the pressure is given by the temperature
dependent part of the free energy density,
\begin{equation}
\label{Pz}
P \, = \, {\varepsilon}_0 - z \, .
\end{equation}
To begin with, let us consider the thermodynamic quantities in the limit of a
zero external field, $H \to 0$. For those quantities we are interested in, we
need all contributions to the pressure which are at most linear in the external
field.

The energy density of the vacuum then reads
\begin{equation}
\label{vacEnergy}
{\varepsilon}_0 \; = \; - {\Sigma}_s H \, + \, {\cal O} (H^2) \, .
\end{equation}
The term quadratic in $H$ involves a logarithm, which depends on a scale
determined by next-to-leading order coupling constants.

The formula for the pressure takes the form
\begin{equation}
\label{PressureH}
P \ = \ \mbox{$ \frac{1}{2}$} (N\!-\!1) g_0 \, + \, 4 \pi a \, (g_1)^2 \, + \,
\pi g \, \Big[ b - \frac{j}{{\pi}^3 F^4} \Big] \, + \, {\cal O} (p^{10}) \, .
\end{equation}
The dependence of the quantity $P$ on temperature is contained in the functions
$g_r(M_{\pi}, T)$ and $j(M_{\pi}, T)$, which are defined in the appendix.

In the limit $H \to 0 \ (\Leftrightarrow T \gg M_{\pi})$ we are interested
in, the functions $g_0, g_1$ and $g$ are given by \cite{Gerber Leutwyler}
\begin{eqnarray}
\label{TempExpg}
g_0 & = & \mbox{$ \frac{1}{45}$} {\pi}^2 T^4 \Bigg[ 1 \, - \,
\frac{15}{4{\pi}^2} \frac{M_{\pi}^2}{T^2} \, + \, {\cal
O}\Big(\frac{M_{\pi}}{T}\Big)^3 \Bigg] \, , \nonumber\\
g_1 & = & \mbox{$ \frac{1}{12}$} T^2 \Bigg[ 1 \, - \,
\frac{3}{\pi} \frac{M_{\pi}}{T} \, + \, {\cal
O}\Big(\frac{M_{\pi}^2}{T^2} \ln\!\frac{M_{\pi}}{T}\Big) \Bigg] \, ,
\nonumber\\
g & = & \mbox{$ \frac{1}{675}$} {\pi}^4 T^8 \Bigg[ 1 \, - \,
\frac{15}{4{\pi}^2} \frac{M_{\pi}^2}{T^2} \, + \, {\cal
O}\Big(\frac{M_{\pi}}{T}\Big)^3 \Bigg] \, ,
\end{eqnarray}
while $j$ diverges logarithmically,
\begin{equation}
j \; = \; \nu \, \ln\!\frac{T}{M_{\pi}} \; + \; j_1 \;
+ \; j_2 \, \frac{M_{\pi}^2}{T^2} \; + \; {\cal O} \Big(\frac{M_{\pi}}{T}
\Big)^3 \, , \quad \nu \equiv \frac{5(N\!-\!1)(N\!-\!2)}{48} \, . 
\end{equation}
The quantities $j_1$ and $j_2$ are real numbers, determined by the group O(N).

The constant $a$ is linear in the external field, whereas $b$ depends
logarithmically on $H$ and involves a scale $H_b$,
\begin{eqnarray}
\label{bConstH}
a & = & - \frac{(N\!-\!1)(N\!-\!3)}{32{\pi}} \frac{{\Sigma}_s H}{F^4} \, ,
\nonumber\\
b & = & - \, \frac{5(N\!-\!1)(N\!-\!2)}{96 {\pi}^3 F^4} \,
\ln\!\frac{H}{H_b} \, .
\end{eqnarray}
The scale $H_b$ is related to the coupling constants $e_1$ and $e_2$ of order
$p^4$ (see appendix).

Equipped with the above formulae, the low temperature expansion of the pressure
amounts to
\begin{equation}
\label{Pressure}
P \ = \ \mbox{$ \frac{1}{90}$} {\pi}^2 (N\!-\!1) \, T^4 \Bigg[ 1 \, + \,
\frac{N\!-\!2}{72} \, \frac{T^4}{F^4} \, \ln{\frac{T_p}{T}} \,
+ \, {\cal O} (T^6) \, \Bigg] \, .
\end{equation}
The first contribution represents the free Bose gas term which originates from
a one-loop graph, whereas the effective interaction among the Goldstone bosons,
remarkably, only manifests itself through a term of order $T^8$. This
contribution contains a logarithm, characteristic of the effective Lagrangian
method, which involves a scale, $T_p$, related to $H_b$ (see appendix).

It is instructive to compare this formula for the pressure with the analogous
relation occurring in theories with a spontaneously broken chiral symmetry,
i.e. for G = $ \mbox{SU(n)}_R \times \mbox{SU(n)}_L \to$ H =
$\mbox{SU(n)}_V\,$ \cite{Leutwyler Uppsala}:
\begin{displaymath}
P \ = \ \mbox{$ \frac{1}{90}$} {\pi}^2 ({n}^2\!-\!1) \, T^4 \Bigg[ 1 \, + \,
\frac{{n}^2    }{144} \, \frac{T^4}{F^4_{\chi}} \, \ln{\frac{T^{\chi}_p}{T}}
\, + \, {\cal O} (T^6) \, \Bigg] \, .
\end{displaymath}
An immediate consistency check of these two results is provided by the
particular case N=4 $\Leftrightarrow$ n=2: since the two groups O(4) and O(3)
are locally isomorphic to SU(2) $\times$ SU(2) and SU(2), respectively, the
above three-loop representations for the pressure have to coincide -- the
formula referring to the O(4) antiferromagnet in zero external field has to be
identical with the one for QCD with two flavors (n=2) in the chiral limit
(zero quark mass). Indeed, this is the case.

The corresponding expressions for the energy density $u$, for the entropy
density $s$ and for the heat capacity $c_V$ are readily worked out from
the thermodynamic relations
\begin{equation}
\label{Thermodynamics}
s \, = \, \frac{{\partial}P}{{\partial}T} \, , \qquad u \, = \, Ts - P \, ,
\qquad c_V \, = \, \frac{{\partial}u}{{\partial}T} \, = \, T \,
\frac{{\partial}s}{{\partial}T} \, ,
\end{equation}
with the result
\begin{eqnarray}
\label{ThermodynQuantities}
u & = & \mbox{$ \frac{1}{30}$} {\pi}^2 (N\!-\!1) \, T^4 \Bigg[ 1 \, + \,
\frac{N\!-\!2}{216} \, \frac{T^4}{F^4} \Big(7 \,
\ln{\frac{T_p}{T}} - 1 \Big) \, + \, {\cal O} (T^6) \, \Bigg]
\, , \nonumber\\
s & = & \mbox{$ \frac{2}{45}$} {\pi}^2 (N\!-\!1) \, T^3 \Bigg[ 1 \, + \,
\frac{N\!-\!2}{288} \, \frac{T^4}{F^4} \Big(8 \, \ln \frac{T_p}{T}
- 1\Big) \, + \, {\cal O} (T^6) \, \Bigg] \, , \nonumber\\
c_V & = & \mbox{$ \frac{2}{15}$} {\pi}^2 (N\!-\!1) \, T^3 \Bigg[ 1 \, +
\, \frac{N\!-\!2}{864} \, \frac{T^4}{F^4} \Big(56 \,
\ln{\frac{T_p}{T}} - 15 \Big) \, + \, {\cal O} (T^6) \, \Bigg] \, .
\end{eqnarray}

The order parameter is given by the logarithmic derivative of the partition
function with respect to the external field,
\begin{equation}
\label{logDer}
{\Sigma}_s (T) \, = \, - \, \frac{{\partial} {\varepsilon}_0}{{\partial} H} \,
+ \, \frac{\partial P}{\partial H} \, .
\end{equation}
This leads to
\begin{eqnarray}
\label{OrdPar}
{\Sigma}_s(T) \ = \ {\Sigma}_s \; \Bigg\{1 \; - \; \frac{N\!-\!1}{24}
\frac{T^2}{F^2} \; 
- \; \frac{(N\!-\!1)\,(N\!-\!3)}{1152} \, \frac{T^4}{F^4} \hspace{3cm}
\nonumber\\
\hspace{3cm} - \; \frac{(N\!-\!1)\,(N\!-\!2)}{1728} \, \frac{T^6}{F^6} \,
\ln{\frac{T_{\Sigma}}{T}} \, + {\cal O} (T^8) \Bigg\} \, .
\end{eqnarray}
The terms of order $T^0, T^2, T^4$ and $T^6$ arise from tree-, one-loop,
two-loop and three-loop graphs, respectively. Up to and including $T^4$, the
coefficients are determined by the constant $F$ which thus sets the scale of
the expansion. The logarithm only shows up at order $T^6$: the scale
$T_{\Sigma}$ involves next-to-leading order coupling constants (see appendix).

As expected, the order parameter gradually melts as the temperature rises. The
effective method, however, has its limitations: the low temperature expansion
can only be trusted at low temperatures -- the curly bracket in (\ref{OrdPar})
represents a correction. In particular, the critical temperature cannot be
accurately determined by setting equation (\ref{OrdPar}) equal to zero.

For nonzero external field, as we have seen before, the low temperature
representations of the thermodynamic quantities and the order parameter retain
their form, except that the coefficients now become functions of $M_{\pi}/T$.
In the region $T \gg {M_{\pi}}$ one recovers the results of the theory for
zero external field, whereas in the opposite limit, $T \ll {M_{\pi}}$, even
the Goldstone bosons freeze. The properties of the system are therefore very
sensitive to the value of the ratio $M_{\pi} / T$. Take e.g. the pressure: in
the limit $H\!\to\!0$, a contribution of order $p^6$, as we have seen, does not
occur. This is no longer the case for an approximate symmetry ($H\!\neq\!0$):
remarkably, the corresponding term of order $p^6$ ($\propto H T^4$) turns out
to be negative (N $\!\neq\,$ 2), signaling an attractive interaction between the
Goldstone degrees of freedom. Note that, with respect to the limit $H\!\to\!0$,
the sign of the effective interaction has changed: there, the first nonleading
term (order $p^8, \, \propto T^8 \ln[T_p/T]$) is positive and the interaction
thus repulsive.

\section{O(3) Antiferromagnet}
\label{AF}

Lorentz invariance is a crucial ingredient of our analysis: in
Lorentz-noninvariant theories, the effective Lagrangian picks up additional
terms. It is therefore not legitimate a priori to transfer the above results to
nonrelativistic systems displaying a spontaneously broken rotation symmetry.

In particular, for N=3, the above low temperature theorems do not in general
hold for systems exhibiting collective magnetic behavior, where the spin waves
or magnons are the relevant Goldstone excitations: in the leading order
effective Lagrangian, a term of topological nature appears, which is
O(3)-invariant only up to a total derivative and violates Lorentz symmetry.
\cite{Leutwyler NRD,Fradkin}. However, this contribution is proportional to the
spontaneous magnetization, such that, for the O(3) {\it anti}\/ferromagnet, it
does not occur -- the leading term of the effective Lagrangian for this system
thus coincides with the leading contribution in the relativistic expression
(\ref{Leff}). Note that, in this analogy, the velocity of light has been
replaced by the spin-wave velocity.

At nonleading order, however, additional terms occur in the effective
Lagrangian, which spoil the formal relativistic invariance. As we have seen in
the preceding section, effective coupling constants of order $p^4$ only show up
in logarithmic scales -- the coefficients in front of the
logarithms exclusively involve leading order coupling constants. Hence, in what
follows, we neglect the complication arising from noninvariant terms of order
$p^4$, and discuss the O(3) antiferromagnet in the framework of the Lagrangian
(\ref{Leff}).

Let us first consider the low temperature expansion of the order parameter --
for the O(3) antiferromagnet, this quantity is referred to as staggered
magnetization. Remarkably, for N=3, the $T^4$-term in formula (\ref{OrdPar})
drops out, and we end up with
\begin{eqnarray}
\label{OrdPar3}
{\Sigma}_s(T) \ = \ {\Sigma}_s \; \Bigg\{1 \; - \; \frac{1}{12}
\frac{(k_{B} T)^2}{ \hbar v  F^2} \; - \; \frac{1}{864}
\frac{(k_{B} T)^6}{{\hbar}^3 v^3 F^6} \,
\ln{\frac{T_{\Sigma}}{T}} \, + {\cal O} (T^8) \Bigg\} \, .
\end{eqnarray}
Note that, for later convenience, we have restored the dimensions: $k_B$
is Boltzmann's constant and $v$ is the spin-wave velocity. The low energy
constant $F$ already appears in the leading $T$-coefficient. By comparing the
above result for the staggered magnetization with the expression derived in
condensed matter physics, we are thus able to identify the effective coupling
constant $F$ in terms of microscopic quantities.

So let us briefly recall how the O(3) antiferromagnet is described within the
microscopic theory. In the Heisenberg model, the exchange Hamiltonian
${\cal H}_0$,
\begin{equation}
\label{HeisenbergModel}
{\cal H}_0 \, = \, - 2 J \sum_{n.n.} {\vec S}_m \cdot {\vec S}_n \, , \qquad J
= const. \, ,
\end{equation}
formulates the dynamics in terms of spin operators ${\vec S}_m$, attached to
lattice sites $m$. Note that the summation only extends over nearest neighbor
pairs and, moreover, the isotropic interaction is assumed to be the same for
any two adjacent lattice sites. If the sign of the exchange integral $J$ is
negative, antiparallel spin alignment is favored, such that we end up with
antiferromagnetic behavior. Clearly, the Hamiltonian is invariant under
rotations of the spin directions, generated by
\begin{equation}
{\vec Q} = \sum_{n} {\vec S}_n \, .
\end{equation}
The ground state of the antiferromagnet, however, does not exhibit this O(3)
symmetry, and its microscopic description is highly nontrivial: in our
analysis, we take it for granted that it spontaneously breaks the symmetry down
to the group O(2).

Moreover, the antiferromagnet is commonly discussed within the following
idealized picture: the system is considered as composed of two sublattices $a$
and $b$, where $a$- and $b$-spins are of equal magnitude and the arrangement is
such that all nearest neighbors of an $a$-spin are $b$-spins and vice versa.
Furthermore, let us assume that the structure of the lattice is simple cubic.
Although the ground state of the antiferromagnet does not exhibit spontaneous
magnetization, the sublattice magnetization itself is not zero. And it is this
latter quantity which is extensively discussed in the literature: indeed, at
leading order in the temperature expansion, a $T^2$-decrease of the sublattice
magnetization has been predicted by many authors
\cite{Kubo,Oguchi,Keffer Loudon,Ludwig,Oguchi Honma},
\begin{equation}
\label{OrdPar3Ludwig}
\frac{M_a(0) - M_a(T)}{g {\mu}_B} \, = \, \frac{1}{2} \frac{V}{a^3} \,
\frac{1}{2{\pi}^2 \sqrt{2z}} \, \Bigg( \frac{k_{B}T}{2|J|S} \Bigg)^2
\zeta(2) \, .
\end{equation}
The expression involves the following quantities: the exchange integral ($J$),
the highest eigenvalue of the spin operator $S^3_n$ ($S$), the number of
nearest neighbors of a given lattice site ($z$), the entire volume of the
system ($V$), the length of the unit cell ($a$), the Land\'e factor ($g$), the
Bohr magneton ($\mu_B$), and the Riemann zeta function.

The sublattice magnetization at zero temperature,
\begin{equation}
\label{MzeroT}
M_a(0) \, = \, \frac{1}{2} \frac{V}{a^3} \, g \mu_B (S - \sigma) \, ,
\end{equation}
involves the quantity $\sigma$, a small number which depends on the structure
of the lattice: in the case of a simple cubic lattice it takes the value
$\sigma = 0.078$ \cite{Anderson 1952}. This relation reflects the well-known
fact that the ground state of the antiferromagnet is highly nontrivial: in
particular, the naive picture where the spin vectors of the two sublattices
point in mutually opposite directions ("N\'eel state"), i.e. $\!\sigma = 0$,
only represents an approximation.

In order to compare the expression (\ref{OrdPar3Ludwig}) referring to the
sublattice magnetization with our result (\ref{OrdPar3}) for the staggered
magnetization, we observe that the two quantities are related via
\begin{equation}
{\Sigma}_s(T) \; = \; \frac{2 M_a(T)}{V} \; = \; \frac{g {\mu}_{B}}{a^3}
\Bigg\{ (S - \sigma) \, - \, \frac{1}{2{\pi}^2 \sqrt{2z}} \,
\Bigg( \frac{k_{B}T}{2|J|S} \Bigg)^2 \zeta(2) + \ldots \Bigg\} \, .
\end{equation}
The above microscopic expression agrees with the effective expansion
(\ref{OrdPar3}) up to order $T^2$, provided that the two coupling constants $F$
and ${\Sigma}_s$ are identified as
\begin{equation}
\label{EffConstMic}
F^2 \; = \; \frac{S - \sigma}{\sqrt{2z}} \, \frac{\hbar v}{a^2}
\; = \; 2S (S - \sigma) \, \frac{|J|}{a} \, , \qquad
{\Sigma}_s \; = \; \frac{g {\mu}_{B}(S - \sigma)}{a^3} \, .
\end{equation}
Note that the spin-wave velocity $v$ is given by the following combination of
microscopic quantities,
\begin{equation}
\label{spinwaveVelocity}
v = \, 2|J|S\sqrt{2z} \, a / \hbar \, .
\end{equation}
The scale of the low temperature expansion is set by $F \sqrt{\hbar v}$ -- let
us briefly estimate its value. Written in terms of the exchange integral $J$,
we obtain
\begin{equation}
F \sqrt{\hbar v} \, = \, 2|J| S \sqrt{(S - \sigma) \sqrt{2z}} \, .
\end{equation}
Now, for a simple cubic lattice ($z \! = \! 6, \sigma \! = \! 0.078$) and for
$S = 1/2$, the double square root on the right hand side is approximately equal
to one, such that we end up with $F \sqrt{\hbar v} \approx |J|$. Typically,
the exchange integral for antiferromagnets is around $|J| \approx 10^{-3} eV$
\cite{Keffer,Rado Suhl 3}, and the scale $F\sqrt{\hbar v}$ thus of the same
order of magnitude. This is to be contrasted with the situation in QCD, where
the relevant quantity, $F_{\chi} \sqrt{\hbar c}$, takes the value $92 MeV$ --
the respective scales in the two theories thus differ in about eleven orders of
magnitude.

As far as subleading terms in the expansion of the staggered magnetization are
concerned, it is well-known that a $T^4$-contribution is absent
\cite{Oguchi,Keffer Loudon,Oguchi Honma}:\cite{footnote5} the spin-wave
interaction only manifests itself at higher orders. Although these authors
predict a term proportional to six powers of the temperature, in agreement with
our result (\ref{OrdPar3}), they do not find the logarithmic dependence on the
temperature. We conclude that it is extremely difficult to calculate the
corrections of order $T^6$ in the framework of a microscopic theory.

As a second comparison, let us now discuss the heat capacity. Setting N=3 in
the effective expansion (\ref{ThermodynQuantities}), we end up with
\begin{equation}
\label{c3}
c_{V} = \mbox{$ \frac{4}{15}$} {\pi}^2 \, \frac{k_{B}^4
T^3}{{\hbar}^3 v^3} \, \Bigg[ 1 \, + \, 
\frac{1}{864} \frac{(k_{B} T)^4}{{\hbar}^2 v^2 F^4} 
\Big(56 \, \ln{\frac{T_p}{T}} - 15 \Big) \, + \, {\cal O} (T^6) \, \Bigg] \, .
\end{equation}
Replacing $v$ according to (\ref{spinwaveVelocity}), the leading contribution
amounts to
\begin{equation}
\label{c3vanKranendonk}
c_{V} = \frac{4 {\pi}^2 \, k_{B}}{15 a^3} \,\Bigg(
\frac{k_{B}T}{2|J|S\sqrt{2z}} \Bigg)^3 \, .
\end{equation}
This expression perfectly agrees with the leading term obtained from a
microscopic or phenomenological analysis of the antiferromagnet
\cite{Kubo,Oguchi,van Kranendonk van Vleck,Akhiezer,Landau Lifshitz}.
 
As far as corrections to the free Bose gas term are concerned, it is also known
that the spin-wave interaction does not manifest itself through a $T^5$-term in
the expansion for the heat capacity \cite{Oguchi,footnote6}. However,
there is again a disagreement with respect to the structure of this
correction: in reference \cite{Oguchi}, a simple $T^7$-contribution is
predicted -- a logarithmic dependence on the temperature is not found.

Nevertheless, it is instructive to compare the two expansions for the heat
capacity at the $T^7$-level. Inserting the microscopic expression
(\ref{EffConstMic}) for $F$ into the effective expansion, we obtain
\begin{displaymath}
c_{V} = \mbox{$ \frac{4}{15}$} {\pi}^2 \, \frac{k_B^4
T^3}{{\hbar}^3 v^3} \, \Bigg[ 1 \, + \, 
\frac{1}{864} \, \frac{2 z \, a^4}{(S\!-\!\sigma)^2}
\frac{(k_{B} T)^4}{{\hbar}^4 v^4} \, \Big(56 \,
\ln{\frac{T_p}{T}} - 15 \Big) \, + \, {\cal O} (T^6) \, \Bigg] \, .
\end{displaymath}
Now, in order for the nonleading term to be consistent with the corresponding
microscopic result of order $T^7$, \cite{Oguchi}
\begin{equation}
\label{OguchiSigmaT^6}
\frac{28 {\pi}^4}{225 \sqrt{3}} \, \frac{1}{S} \, k_{B} \,
\frac{a^4}{(\hbar v)^7} \, (k_{B} T)^7 \, ,
\end{equation}
the logarithm must have the following numerical value ($S = 1/2$, simple cubic
lattice): $\ln(T_p / T) \approx 1.5$. Hence, for that specific value of the
temperature, $T_0 = 0.23 \, T_p$, the two results coincide.

Let us consider the analogous situation in QCD: there, the value of
$T_p^{\chi}$ can be extracted from experiment ($\pi\pi$-scattering, for
details see e.g.\cite{Gerber Leutwyler}), yielding $T_p^{\chi} \approx 275 \,
MeV$. Accordingly, with the above value of the logarithm, we get $T^{\chi}_0 
\approx 60 MeV$. The critical temperature is estimated to be around $170 MeV$
-- so we see that the values for $T^{\chi}_0, T^{\chi}_c$ and $T^{\chi}_p$ are 
of the same order of magnitude. Note that the respective scales for the 
antiferromagnet differ in about ten orders of magnitude: with a critical 
temperature (N\'eel temperature) of $T_N \approx {\cal O}(0.01 \, eV)$, we end 
up with $T_p \approx {\cal O}(0.01 \, eV)$.
 
\section{Summary and Outlook}
\label{SumOut}

The presence of states with small excitation energies affects the behavior of
the system in a very specific manner, controlled by the symmetries of the
underlying theory. These symmetries unambiguously fix the values of the
coefficients in the low temperature expansion of the order parameter and the
thermodynamic quantities up to two leading order coupling constants, $F$ and
${\Sigma}_s$. Symmetry, however, does not determine the logarithmic scales
$T_p$ and $T_{\Sigma}$, which occur in the temperature expansion and involve
next-to-leading order coupling constants.

The low temperature theorems for the order parameter and the thermodynamic
quantities are exact up to and including three loops: independently of the
specific underlying model, they are valid for any Lorentz-invariant theory
where an O(N) symmetry is spontaneously broken to O(N--1). For N=4, the
expansion (\ref{OrdPar}) for the order parameter not only describes the quark
condensate of QCD with two flavors in the chiral limit, but also, e.g.,
describes the Higgs condensate in the Standard Model of elementary particle
physics. The only difference between the two representations concerns the
numerical value of the coupling constants occurring therein -- the low
temperature description turns out to be universal.

Another interesting case, which we have discussed in detail, is given by N=3:
the O(3) antiferromagnet. The low temperature theorems for the staggered
magnetization and the heat capacity both agree with the results given in the
literature up to and including two loops. At the three-loop level, however, the
results no longer coincide: to my knowledge, the logarithmic temperature
dependence at order $T^7$ in the expansion for the heat capacity and order
$T^6$ for the staggered magnetization, has been overlooked so far. The
effective Lagrangian method not only proves to be more efficient than the
complicated microscopic analysis, but also addresses the problem from a unified
point of view based on symmetry -- at large wavelengths, the microscopic
structure of the system only manifests itself in the numerical values of a few
coupling constants.

Although Lorentz invariance is a crucial ingredient of the present analysis,
the effective Lagrangian method is not restricted to this domain: ferromagnets,
e.g., where the spin waves follow a quadratic dispersion law, may be analyzed
within the framework of nonrelativistic effective Lagrangians \cite{Hofmann
Scattering,Leutwyler NRD,Soto}. In particular, the low temperature expansion
for the order parameter of a ferromagnet, its spontaneous magnetization, has
been calculated to three loops -- the results, which go beyond Dyson's
pioneering microscopic analysis \cite{Dyson}, will be presented in a
forthcoming paper \cite{Hofmann Ferro}.

\acknowledgements
I would like to thank H. Leutwyler for his patient assistance throughout this
work and for his critical reading of the manuscript. Thanks also to G.
Colangelo, S. Mallik and D. Toublan for their help. I am greatly indebted to
the Janggen-P\"ohn-Stiftung for supporting my doctoral thesis. Likewise,
support by Schweizerischer Nationalfonds is gratefully acknowledged.

\section*{Appendix}
\renewcommand{\theequation}{A.\arabic{equation}}
\setcounter{equation}{0}
In this appendix, we discuss the various quantities showing up in the formula
for the pressure,
\begin{equation}
\label{PressureFull}
P \ = \ \mbox{$ \frac{1}{2}$} (N\!-\!1) g_0 \, + \, 4 \pi a \, (g_1)^2 \, +
\, \pi b g \, - \, \frac{1}{F^4} I \, + \, {\cal O}(p^{10}) \, .
\end{equation}
Unlike in section 2, we do not restrict ourselves to contributions at most
linear in the external field. In the second part of the appendix, we briefly
comment on the logarithmic scales $T_p$ and $T_{\Sigma}$.

Let us first consider the renormalized Goldstone-boson mass. The calculation
yields
\begin{equation}
\label{renMass}
M_{\pi}^2 \ = \ \frac{{\Sigma}_s H}{F^2} \, + \, [ \, 2 \, (k_2 - k_1) \, +
\, (N\!-\!3) \, \lambda ] \frac{({\Sigma}_s H)^2}{F^6} \; + \; c \,
\frac{({\Sigma}_s H)^3}{F^{10}} \; + \; {\cal O} (H^4) \, ,
\end{equation}
where the constant $c$ involves the relevant coupling constants of order $p^6$.
The quantity $\lambda$ contains a pole at $d\!=\!4$,
\begin{eqnarray}
\label{lambda}
\lambda & = & \mbox{$ \frac{1}{2}$} \, (4 \pi)^{-d/2} \;
\Gamma(1-{\mbox{$ \frac{1}{2}$}}d) \; M^{d-4} \nonumber\\  & = &
\frac{M^{d-4}}{16{\pi}^2} \, \Bigg[ \frac{1}{d-4} \, - \, \mbox{$
\frac{1}{2}$} \{ \ln{4{\pi}} + {\Gamma}'(1) + 1 \} \, + \, {\cal O} (d\!-\!4)
\Bigg] \, .
\end{eqnarray}
This singularity is absorbed in a renormalization of the combination
$k_2\!-\!k_1$ of coupling constants of order $p^4$. One ends up with a
logarithm depending on $H$ and a term independent thereof. The latter can be
absorbed in a new scale $H_{M}$, such that the expression for the
renormalized mass takes the form displayed in (\ref{renMassH}),
\begin{equation}
\label{renMassH2}
\hspace{2.8cm} M_{\pi}^2 \ = \ \frac{{\Sigma}_s H}{F^2} \, + \,
\frac{N\!-\!3}{32{\pi}^2} \, \frac{({\Sigma}_s H)^2}{F^6} 
\ln\frac{H}{H_{M}}
\; + \; {\cal O} (H^3) \, .
\end{equation}

Now, the functions $g_0, g_1, g$ and $I$ occurring in the formula
(\ref{PressureFull}) for the pressure, depend in a nontrivial manner on
$M_{\pi}$ and $T$. The quantities $g_r$ are associated with the
$d$-dimensional noninteracting Bose gas,
\begin{equation}
\label{FreeFunctions}
g_r(M_{\pi}, T) \, = \, 2 {\int}_{\!\!\! 0}^{\infty} \frac{\mbox{d} \rho}{(4
\pi \rho)^{d/2}} \, {\rho}^{r-1} \, \exp(- \rho M^2_{\pi})
\sum_{n=1}^{\infty} \exp(- n^2/{4 \rho T^2}) \, .
\end{equation}
The function $g$ is the following combination thereof,
\begin{equation}
\label{gCombSimpl}
g \, = \, 3 g_0 \, (g_0 + M^2_{\pi} \, g_1) \, .
\end{equation}
The expression for the three-loop integral $I$ is more complicated:
\begin{eqnarray}
\label{FunctionI}
I \ = \ \mbox{$ \frac{1}{48}$} (N\!-\!1)(N\!-\!3)M^4_{\pi} {\bar J}_1 \; -
\; \mbox{$ \frac{1}{4}$} (N\!-\!1)(N\!-\!2) {\bar J}_2 \hspace{4cm}
\nonumber\\
\; - \; \mbox{$ \frac{1}{16}$} (N\!-\!1)(N\!-\!3)^2 M^4_{\pi} \, (g_1)^2 g_2 +
\; \mbox{$ \frac{1}{48}$} \, (N\!-\!1)(N\!-\!3)(3N\!-\!7) M^2_{\pi} \,
(g_1)^3 \, .
\end{eqnarray}
The quantities ${\bar J}_1$ and ${\bar J}_2$,
\begin{eqnarray}
\label{FunctionsJbarJ}
{\bar J}_1 & = & J_1 \, - \, c_1 \, - \, c_2 \!\, g_1 \, + \, 6
(d\!-\!2){\lambda} (g_1)^2 \, , \nonumber\\
{\bar J}_2 & = & J_2 \, - \, c_3 \, - \, c_4 \!\, g_1 \, + \, \mbox{$
\frac{1}{3}$} (d\!+\!6)(d\!-\!2) {\lambda} \Big({\bar G}_{\mu \nu}\Big)^2 \,
+ \, \mbox{$ \frac{2}{3}$} (d\!-\!2) {\lambda} M^4_{\pi} \, (g_1)^2 \, ,
\nonumber\\
{\bar G}_{\mu \nu} & = & - \mbox{$ \frac{1}{2}$} {\delta}_{\mu \nu} g_0
\, + \, {\delta}^4_{\mu} \, {\delta}^4_{\nu} \, ( \mbox{$ \frac{1}{2}$} d g_0
+ M^2_{\pi} g_1) \, ,
\end{eqnarray}
remove the singularities of the loop integrals $J_1$ and $J_2$, respectively:
\begin{eqnarray}
\label{FunctionsJ}
J_1 & = & {\int}_{\!\!\!T} \! {\mbox{d}}^dx \; [G(x)]^4 \, , \nonumber\\
J_2 & = & {\int}_{\!\!\!T} \! {\mbox{d}}^dx \; [{\partial}_{\mu} G(x)
{\partial}_{\mu} G(x)]^2 \, .
\end{eqnarray}
For the explicit structure of the temperature independent counterterms
$c_1 \dots c_4$, the reader is referred to reference \cite{Gerber Leutwyler}.

The connection between the formula (\ref{PressureH}) for the pressure given in
section 2,
\begin{displaymath}
P \ = \ \mbox{$ \frac{1}{2}$} (N\!-\!1) g_0 \, + \, 4 \pi a \,(g_1)^2 \, + \,
\pi g \, \Big[b - \frac{j}{{\pi}^3 F^4}\Big] \, + \, {\cal O} (p^{10}) \, ,
\end{displaymath}
and (\ref{PressureFull}) is established by splitting off a factor $g$ from the
expression $I$,
\begin{equation}
I \, = \, \frac{1}{{\pi}^2} \, g \, j \, .
\end{equation}
This relation defines the function $j$.

The constants $a$ and $b$ in (\ref{PressureFull}) contain the various coupling
constants which occur in the effective Lagrangian,\cite{footnote7}
\begin{eqnarray}
\label{aConst}
a \ = \ - \frac{(N\!-\!1)(N\!-\!3)}{32{\pi}} \frac{{\Sigma}_s H}{F^4} \, +
\, \frac{N\!-\!1}{4{\pi}} \frac{({\Sigma}_s H)^2}{F^8} \, \Bigg\{
\Big[(N\!+\!1)(e_1 + e_2) + k_2 - k_1 \Big] \nonumber\\
 - \, \frac{(N\!-\!1)^2}{2}
\lambda \, - \,\frac{3{N}^2 + 32N - 67}{768\,{\pi}^2} \Bigg\} \, ,
\nonumber\\ 
b \ = \ \frac{N\!-\!1}{{\pi}F^4} \, \Bigg\{ \Big[ 2 e_1 + N e_2 \Big] \, -
\, \frac{5(N\!-\!2)}{3} \lambda \, - \, \frac{N\!-\!2}{16{\pi}^2} \Bigg\}
\, . \hspace{3cm}
\end{eqnarray}
Repeating the steps which led from (\ref{renMass}) to (\ref{renMassH2}), the
constants $a$ and $b$ may be conveniently written as
\begin{eqnarray}
\label{bConstLog}
a & = & - \, \frac{(N\!-\!1)(N\!-\!3)}{32{\pi}} \frac{{\Sigma}_s H}{F^4} \,
- \, \frac{(N\!-\!1)^3}{256{\pi}^3} \frac{({\Sigma}_s H)^2}{F^8}
\ln\frac{H}{H_a} \, , \nonumber\\
 b & = & - \frac{5(N\!-\!1)(N\!-\!2)}{96{\pi}^3 F^4} \, \ln\frac{H}{H_b} \, .
\end{eqnarray}

Finally, let us consider the logarithmic scales $T_p$ and $T_{\Sigma}$,
showing up in the low temperature expansion of the thermodynamic quantities and
the order parameter. They are both related to the scale ${\Lambda}_b$,
\begin{equation}
T_p = {\Lambda}_b \, \exp(- j_1/\nu) \, , \quad
T_{\Sigma} = {\Lambda}_b \, \exp\Big(- j_1/\nu + \frac{4{\pi}^2}{15 \nu} j_2
\Big) \, ,
\end{equation}
where
\begin{equation}
{\Lambda}_b = \frac{{\sqrt{{\Sigma}_s H_b}}}{F} \, , \qquad \nu \equiv
\frac{5(N\!-\!1)(N\!-\!2)}{48} \, .
\end{equation}
Note that $T_p$ and $T_{\Sigma}$ depend on the constants $j_1$ and $j_2$
occurring in the expansion of the function $j$ (in the limit $T \gg
M_{\pi}$),
\begin{equation}
\label{jExpansion}
j \; = \; \nu \, \ln \frac{T}{M_{\pi}} \; + \;
j_1 \; + \; j_2 \, \frac{M^2_{\pi}}{T^2} \; + \; {\cal O}
\Big(\frac{M_{\pi}}{T}\Big)^3 \, .
\end{equation}

The formulae given in this appendix can readily be checked by setting N=4:
using $M^2 = {\Sigma}_s H / F^2$ and identifying the respective coupling
constants as $e_1\!=\!l_1, \, e_2\!=\!l_2$ and $k_2\!-\!k_1\!=\!l_3$, the
results for two-flavor QCD are reproduced \cite{Gerber Leutwyler,Hasenfratz
Leutwyler}.


\begin{references}
\bibitem{footnote1}
Reference \cite{Burgess} also addresses the issues of applying techniques
developed for particle physics to condensed matter systems with Goldstone
bosons. A specific application concerning SO(5) invariance and high-$T_c$
superconductivity can be found in \cite{Burgess Lutken}.
\bibitem{footnote2}
An example related to condensed matter physics concerns systems exhibiting
collective magnetic behavior. Ferro- and antiferromagnets are described within
the O(3)-invariant Heisenberg model. Interactions which explicitly break the
rotation symmetry, such as external magnetic and anisotropy fields
(see \cite{Hofmann Scattering}) or magnetic dipole and spin-orbit interactions
(see \cite{Soto}), can easily be incorporated into the effective Lagrangian
machinery.
\bibitem{footnote4}
In quantum chromodynamics this scale is given by ${\Lambda}_{QCD}$, the
renormalization group invariant scale of QCD; in connection with
antiferromagnets, the relevant scale may be identified with the exchange
integral $J$ in the Heisenberg model.
\bibitem{footnote5}
If the discreteness of the lattice is taken into account, one ends up with
corrections proportional to $T^4, T^6, \dots$ \cite{Oguchi}.
\bibitem{footnote6}
We are not considering effects due to the discreteness of the lattice, which
yield additional terms proportional to $T^5, T^7, \dots$ \cite{Oguchi}.
\bibitem{footnote7}
Note that, with respect to the last terms in the brace ($\propto {\pi}^{-2}$),
our definition differs from the convention used in reference
\cite{Gerber Leutwyler}: the authors split these terms between (\ref{aConst})
and the function $I$ (\ref{FunctionI}), such that, for N=4 (strong interaction)
the constants $a$ and $b$ may be identified with sums of scattering lengths and
effective ranges (pion-pion scattering).
\end{references}
\end{document}